# Usage of OpenAlex for creating meaningful global overlay maps of science on the individual and institutional levels

Robin Haunschild[*] and Lutz Bornmann[*, **]

[*]R.Haunschild@fkf.mpg.de (0000-0001-7025-7256), L.Bornmann@fkf.mpg.de (0000-0003-0810-7091)
Information Retrieval Service, Max Planck Institute for Solid State Research, Germany

[**]bornmann@gv.mpg.de
Science Policy and Strategy Department, Administrative Headquarters of the Max Planck Society, Germany

**Abstract**
Global overlay maps of science use base maps that are overlaid by specific data (from single researchers, institutions, or countries) for visualizing scientific performance such as field-specific paper output. A procedure to create global overlay maps using OpenAlex is proposed. Six different global base maps are provided. Using one of these base maps, example overlay maps for one individual (the first author of this paper) and his research institution are shown and analyzed. A method for normalizing the overlay data is proposed. Overlay maps using raw overlay data display general concepts more pronounced than their counterparts using normalized overlay data. Advantages and limitations of the proposed overlay approach are discussed.

## 1. Introduction

The popularity of science maps in empirical bibliometrics today has settled into theorizing on bibliometrics. Science maps conceptualize science as a communication system, where "information obtained at certain nodes is transmitted to other nodes" (Ziman, 2000, p. 113). Citations play an important role in these communication systems, since they form the bridges between published research (Roth, 2005) and may point out research communities and their connections. Tahamtan and Bornmann (2022) recently introduced a new citation theory that is based on the social systems theory proposed by Niklas Luhmann (2012a, 2012b). The Social Systems Citations Theory (SSCT) focusses on communications in science which are formally manifested as publications and citations in scientific communication networks. These networks can be observed and empirically studied by using science maps accompanied by the SSCT: Science maps are then conceptually rooted in the SSCT (Sjögårde, 2023; Tahamtan & Bornmann, 2022).

This study deals with a specific form of science maps: global science overlay maps. These maps apply base maps (which usually include the whole science system), on which certain overlay data are presented. Overlay maps visualize how the overlaid data (e.g., research of a researcher or an institution) are positioned in the whole science system. We introduce in this study the use of OpenAlex (Priem, Piwowar, & Orr, 2022) data (a new bibliometric database which is freely available at https://openalex.org) to build meaningful global science maps which can be overlaid by the user with own data. For example, specific institutional overlay data can be used in combination with the base maps from this study to explore the field-specific activities of the institution.

## 2. Global science overlay maps

The visualization of data as overlay maps is a development in bibliometrics that started around a decade ago (Leydesdorff, Rafols, & Chen, 2013). The generation of global overlay maps include two steps: In the first step, meaningful global base maps are developed which can be used as bases for overlays with specific data. Meaningful global base maps position certain elements (e.g., publications or fields) in such a way that most of the elements are well visible to an observer. In the second step, the data of specific interest are downloaded from databases and overlaid on the global base map. Overlay maps have been widely used in different contexts in recent years. For example, some authors have applied the technique to overlay bibliometric data on Google maps (Bornmann & Leydesdorff, 2011; Bornmann, Leydesdorff, Walch-Solimena, & Ettl, 2011; Leydesdorff & Persson, 2010). The global overlay mapping technique was initially introduced by Boyack (2009). Klavans and Boyack (2009) were the first to propose a global map of science that can be used as base map. The technique was further developed, e.g., by Rafols, Porter, and Leydesdorff (2010), Leydesdorff and Rafols (2012), and Leydesdorff, et al. (2013).

The popular VOSviewer software (van Eck & Waltman, 2010) can be used to produce meaningful and aesthetic global overlay maps. Having large amounts of publication data, the construction of a global base map is always time consuming and resource demanding in contrast to the overlay procedure. We are not aware of any openly and freely accessible global base map based on any multi-disciplinary bibliometric database which can be used to overlay own data. Thus, in this contribution, we aim to provide openly and freely accessible global base maps using all publications indexed in OpenAlex that can be used by anyone to overlay particular data of interest (e.g., publication sets of single institutions or researchers). Since OpenAlex is a free and open catalog of the global research system, it can be used by everyone without restriction for collecting the interesting overlay data.

## 3. Methods

### *3.1. Datasets*

Currently, OpenAlex (Priem, et al., 2022) provides snapshots once a month. We used the snapshot from August 2023 for our analyses as available from the Competence Center for Bibliometrics, CCB, see https://bibliometrie.info/). The snapshot contains 243,053,925 documents. We used the following different time periods for our analyses: (i) 1800-2022 with 237,876,541 documents, (ii) 2008-2022 with 134,092,007, (iii) 2013-2022 with 95,438,459 documents, (iv) 2018-2022 with 47,665,990 documents, and (v) 2022 with 8,496,167. We constructed global maps of science using these time periods. Depending on the publication years of the overlay data, the user can select one of these base maps to construct the global overlay map. We used concepts algorithmically assigned by OpenAlex to documents as nodes on the map. OpenAlex provides concepts on six different levels (from 0 to 5). According to the OpenAlex website, "concepts are abstract ideas that works are about" (https://docs.openalex.org/api-entities/concepts).

We explored different possibilities of using each level separately or combining concepts of different levels to produce well-suited base maps. The maps based on concepts of the top three levels (i.e., 0, 1, and 2) produced the best maps to have meaningful insights into worldwide research. Thus, we only include documents that have a concept of level 0, 1, or 2 assigned to them. As we want to provide global maps covering all research, we do not impose any restrictions on document types.

### *3.1. Statistics*

For calculating the positions of the concepts on the base maps, we used direct citation relations between the documents that belong to these concepts. For all maps but one, we used a citation window of five years plus the publication year. We produced one version of the map that includes the time period 1800-2022 with a citation window of 30 years plus the publication year. We calculated the citation relations as reference relations so that the citation window can be equally long for each document, i.e., citation relations between 1992 (2017) and 2022 were considered for documents published in 2022, and citation relations between 1770 (1795) and 1800 were considered for documents published in 1800 for a citation window of 30 (5) years.

The global base maps were overlaid with two different datasets in this study to demonstrate the overlay technique using OpenAlex data: one for a researcher and another for an institution.

Both are well interpretable for us: (i) documents assigned to the OpenAlex author ID of the first author of this paper (RH) and (ii) documents assigned to the OpenAlex institution ID of the Max Planck Institute for Solid State Research (MPI-FKF). We present and discuss the example maps with respect to advantages and disadvantages of the overlay technique. We kept the VOSviewer parameters label size variation of 0.2 and scale of 0.5, for the map of RH. To have a better distinction between the node sizes for the institutional map, we used size variation of 0.4 and scale of 1.0 for the MPI-FKF map. We recommend that the optimal values for size variation and scale are individually chosen for the overlay maps by the users. However, two overlay maps that are compared with one another should be created using the same parameters. Data to overlay on the global base maps can be exported from the OpenAlex web-interface, downloaded from the OpenAlex API, or exported from an OpenAlex snapshot. Readers can download the global base maps and the overlay maps discussed in this paper (Haunschild & Bornmann, 2024) at [http://ivs.fkf.mpg.de/global_maps_OpenAlex](http://ivs.fkf.mpg.de/global_maps_OpenAlex). Thus, readers can experiment with other VOSviewer parameters using our example overlay maps.

For the comparison of overlay maps, it is necessary to apply normalization procedures to the overlay data. We propose to normalize the overlay data by dividing the proportion of documents per concept in the focus dataset by the proportion of documents per concept in OpenAlex:

$$N_{Wl} = \sum_{c} N_{cWl}$$

$$N_{Ul} = \sum_{c} N_{cUl}$$

$$p_{cWl} = {N_{cWl}}/{N_{Wl}}$$

$$p_{cUl} = {N_{cUl}}/{N_{Ul}}$$

$$a_{cUl} = {p_{cUl}}/{p_{cWl}}$$

Here, $N_{Wl}$ and $N_{Ul}$ are the total number of documents of the world ($W$) and the overlaid focal unit ($U$) on the same level $l$ as the concept $c$. The number of documents assigned to specific concepts ($c$) of the world and the focal unit are $N_{cWl}$ and $N_{cUl}$. The proportions of documents assigned to specific concepts of the world and the focal unit are $p_{cWl}$ and $p_{cUl}$. The activity, i.e., normalized proportion of documents assigned to specific concepts of the focal unit is $a_{cUl}$. These activities replace the $N_{cUl}$ values in the column "weight<papers>". This normalization procedure calculates the ratio of the proportion of documents assigned to a certain concept of the focal unit to the world's proportion of papers assigned to the same concept. This means that equally sized nodes on the raw maps will be of different sizes on the normalized maps if the world's proportions for these concepts are different. Global reference data needed for the normalization procedure and overlay data for authors and institutions can be retrieved from the OpenAlex API.

## 4. Results

In section 4.1, the base maps are explained that can be used for overlay maps. The examples in section 4.2 demonstrate how certain overlays can be applied to the base maps and how the maps can be interpreted.

### *4.1. Base maps*

Figure 1 shows one of the six global base maps that we produced in this study with the main scientific fields additionally annotated. The clusters were determined by the VOSviewer algorithm for association strength using citation relations. This gave rise to six different clusters. The cluster colors were assigned by the cluster size: The largest cluster is colored in orange, the second largest in green, the third largest in blue, the fourth largest in yellow, the fifth largest in purple, and the smallest cluster in light blue. The other base maps are available at [http://ivs.fkf.mpg.de/global_maps_OpenAlex](http://ivs.fkf.mpg.de/global_maps_OpenAlex) (Haunschild & Bornmann, 2024).The orange cluster in Figure 1 contains 6,788 concepts mainly from the social sciences and the humanities. The green cluster mainly contains 4,210 medicinal concepts. The blue cluster contains 3,247 concepts from the fields of physics and engineering. The yellow cluster contains 3,153 concepts from the fields of mathematics, computer sciences, and theoretical physics. The purple cluster contains 2,825 concepts mainly related to biology. The light blue cluster covers 1,534 concepts from chemistry and material sciences.

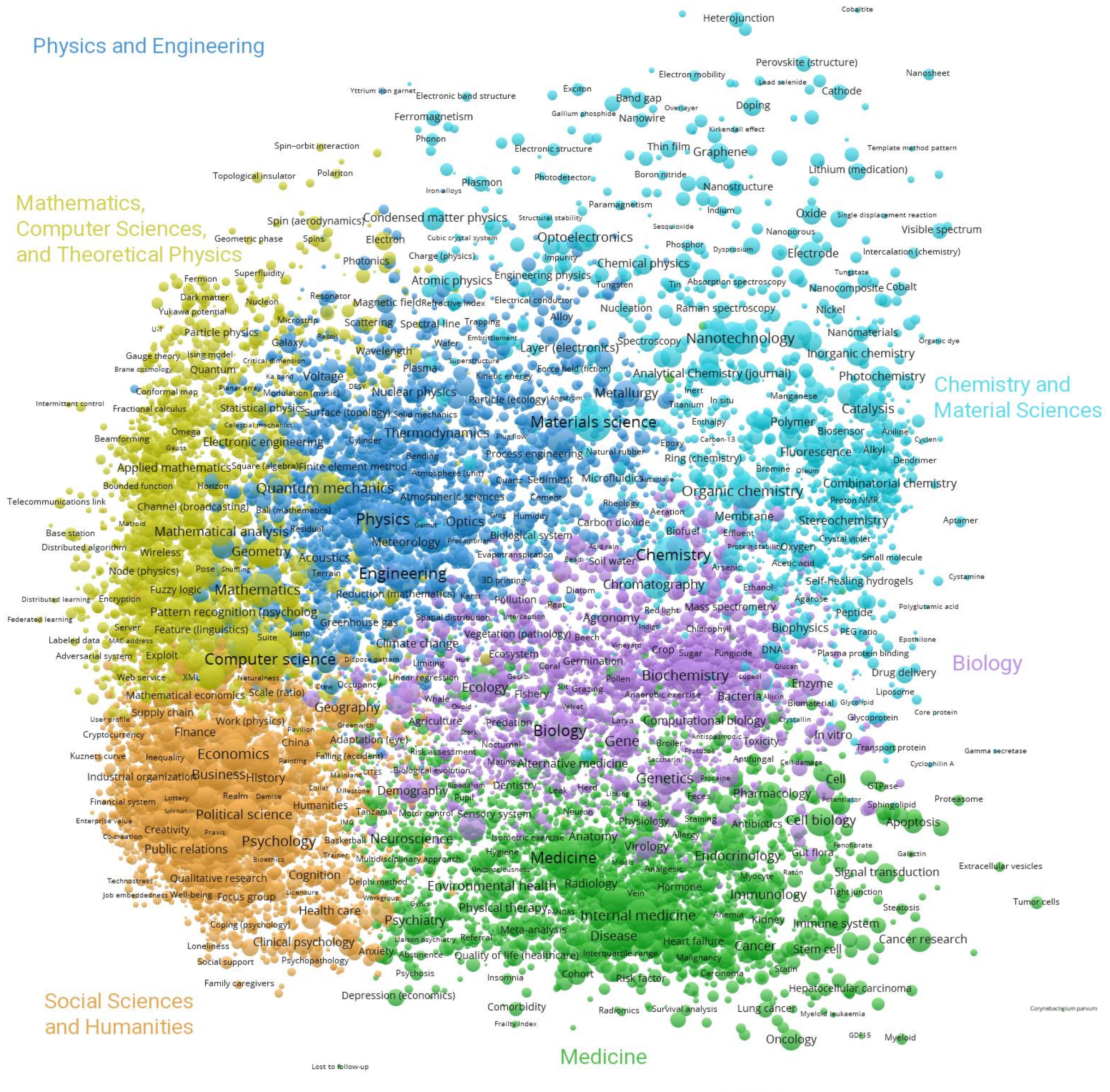

Figure 1. Global base map using the time period 2008-2022 with the main scientific fields additionally annotated

## *4.2. Two examples for overlay maps*

In this section, we show two overlay maps as examples by using the base map in Figure 1. On the one hand, the example maps are intended to demonstrate the usefulness of the overlay approach. On the other hand, we discuss based on the specific overlay maps whether the overlay approach using OpenAlex data produces reliable and meaningful results. We generated overlay maps that can be well interpreted by us. The approach developed in this study stands or falls with the quality of the concepts provided by OpenAlex, since the most important information on the overlaid research are the labels from the concepts. Figure 2 and Figure 3 show the overlay maps for two different focus datasets covering a researcher and an institution, respectively.

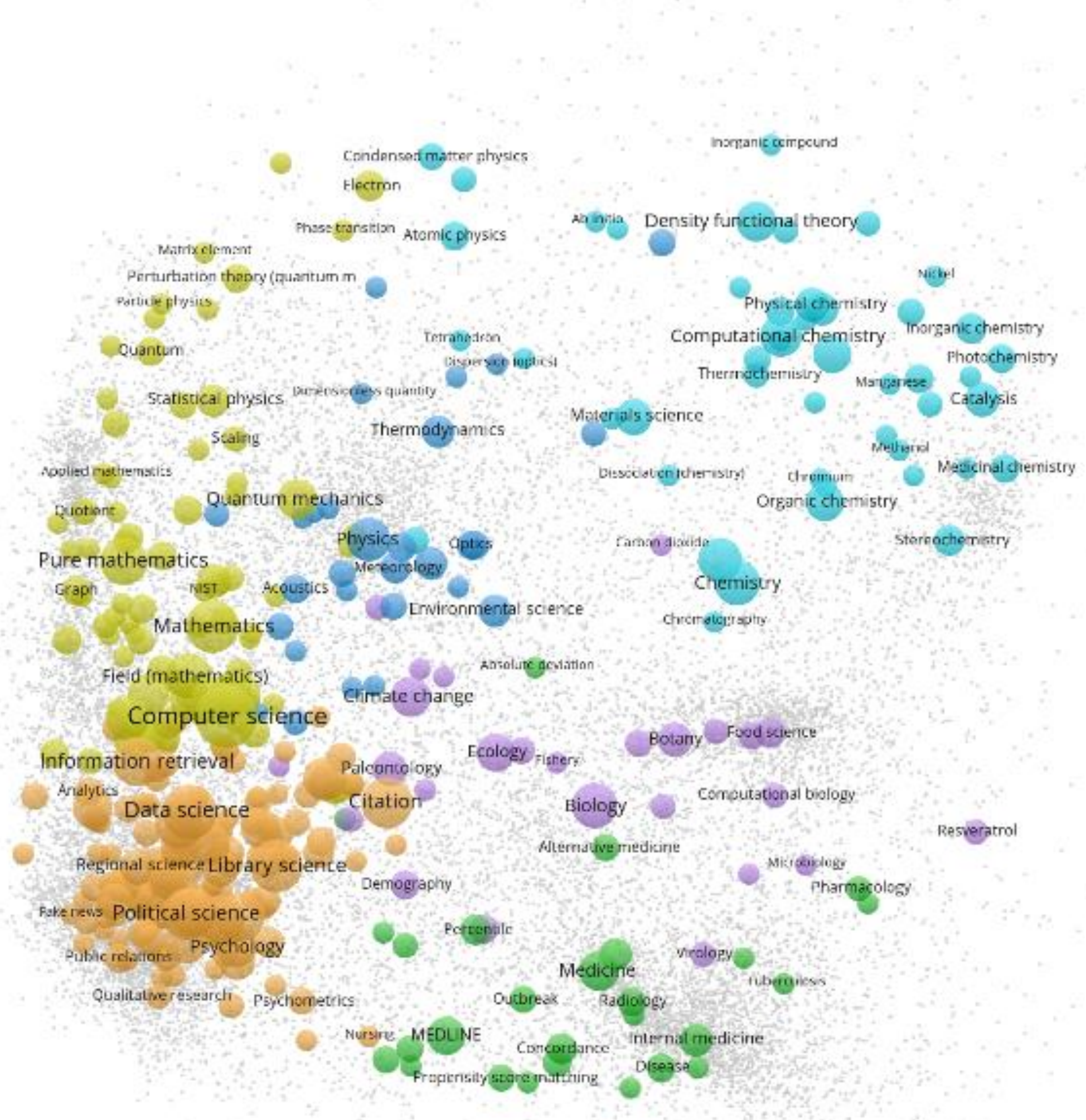

Figure 2. Global overlay map for documents assigned to the OpenAlex author ID of the first author of this paper (RH)

Before conducting research in bibliometrics, RH (see https://www.researchgate.net/profile/Robin_Haunschild) has done research in computational and theoretical chemistry (see Figure 2). This is reflected in the pronounced light blue nodes, e.g., Computational chemistry, Density functional theory, and Catalysis, and some of the yellow nodes, e.g., Perturbation theory (quantum mechanics). RH's current research activities can be seen in the pronounced orange cluster, e.g., Data science, Citation, Library science, and Information retrieval. However, some bibliometric papers seem to also be assigned to concepts such as Computer science, Mathematics, and Field (mathematics). The latter most probably is related to the area of field normalization in bibliometrics. Without insight into the dataset, the explanatory addition "mathematics" in parentheses is more confusing than helpful. That many documents are assigned to Computer Science may be explained by the fact that most of RH's journal papers are published in the journal *Scientometrics*. 88.9% of the *Scientometrics* papers are assigned to the concept Computer science. Since about half of the papers published in the journal *Scientometrics* is assigned to the concepts Political science and Library science, there is also a pronounced node Political science in Figure 2. The green and purple nodes are partly due to bibliometric papers about environmental, medical, and biological fields, e.g., Climate change, Biology, and Medicine. This also holds true for some of the blue nodes, e.g., Meteorology. RH has checked the 30 documents with DOI assigned to his author ID and the concept Biology in OpenAlex.[1] The documents were correctly assigned to RH. However, the assignments of most documents to Biology are at best questionable.

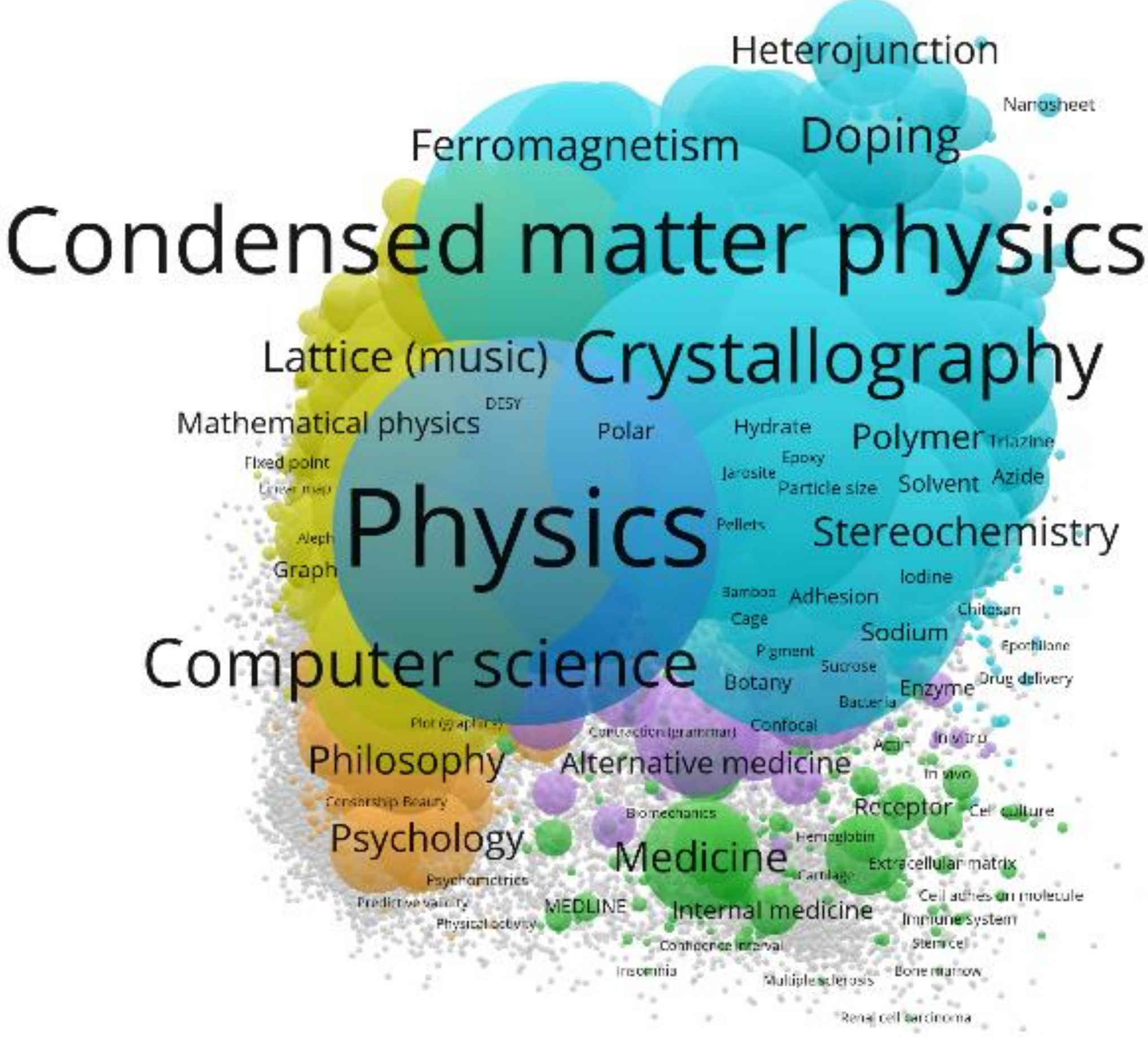

Figure 3. Global overlay map for documents assigned to the OpenAlex institution ID of the Max Planck Institute for Solid State Research (MPI-FKF)

Since the MPI-FKF (see https://www.fkf.mpg.de/en) is broadly covering chemistry and physics of condensed matter from a theoretical and an experimental perspective, the main nodes in Figure 3 appear in the yellow (mathematics, computer sciences, and theoretical physics), blue (physics and engineering), and light blue (chemistry and material sciences) clusters. The largest concepts, e.g., Physics, Condensed matter physics, and Crystallography, clearly represent some of the main research activities of the MPI-FKF. Like in the map for RH, we also see in the MPI-

---

1 https://openalex.org/works?sort=cited_by_count%3Adesc&column=display_name,publication_year,type,open_access.is_oa,cited_by_count&page=1&filter=authorships.author.id%3AA5004115158,concepts.id%3AC86803240

FKF map concepts with confusing additions in parentheses, e.g., Lattice (Music). Most of the MPI-FKF papers assigned to this concept are about crystal structure research by retired directors (e.g., Manuel Cardona, Arndt Simon, Martin Jansen, and Joachim Maier). The nodes in the other clusters are somewhat surprising. Some of the nodes in the orange, green, and purple clusters might be due to the research activities of RH and his predecessor Werner Marx. For example, the typical nodes for bibliometrics, such as MEDLINE and Citation (covered by other nodes in Figure 3), appear not only in Figure 3, but also on the overlay map of RH (Figure 2). Concepts from biomedicine are emphasized too much on the map of MPI-FKF. Since some nodes, e.g., In vivo and Immune system, do not appear on the overlay map of RH's documents, they do not seem to connect to his research. Such nodes probably indicate a problematic assignment of concepts to documents in OpenAlex.

Figure 4 and Figure 5 show the normalized versions of the global overlay maps from Figure 2 and Figure 3. It seems that the normalized versions of the maps provide more meaningful insight into the units' research than the versions without normalization. Comparing non-normalized and normalized figures, we see in general that some concepts are more pronounced in the normalized overlay version than in the raw overlay versions. For example in Figure 4, the concepts Scientometrics, Bibliometrics, and Topic model are much more pronounced than in Figure 2. Similar observations can be made for the MPI: The concepts MEDLINE, Medicine, and Internal medicine are much smaller in Figure 5 than in Figure 3. More general concepts are more pronounced on the raw maps whereas more specific concepts are more pronounced on the normalized maps.

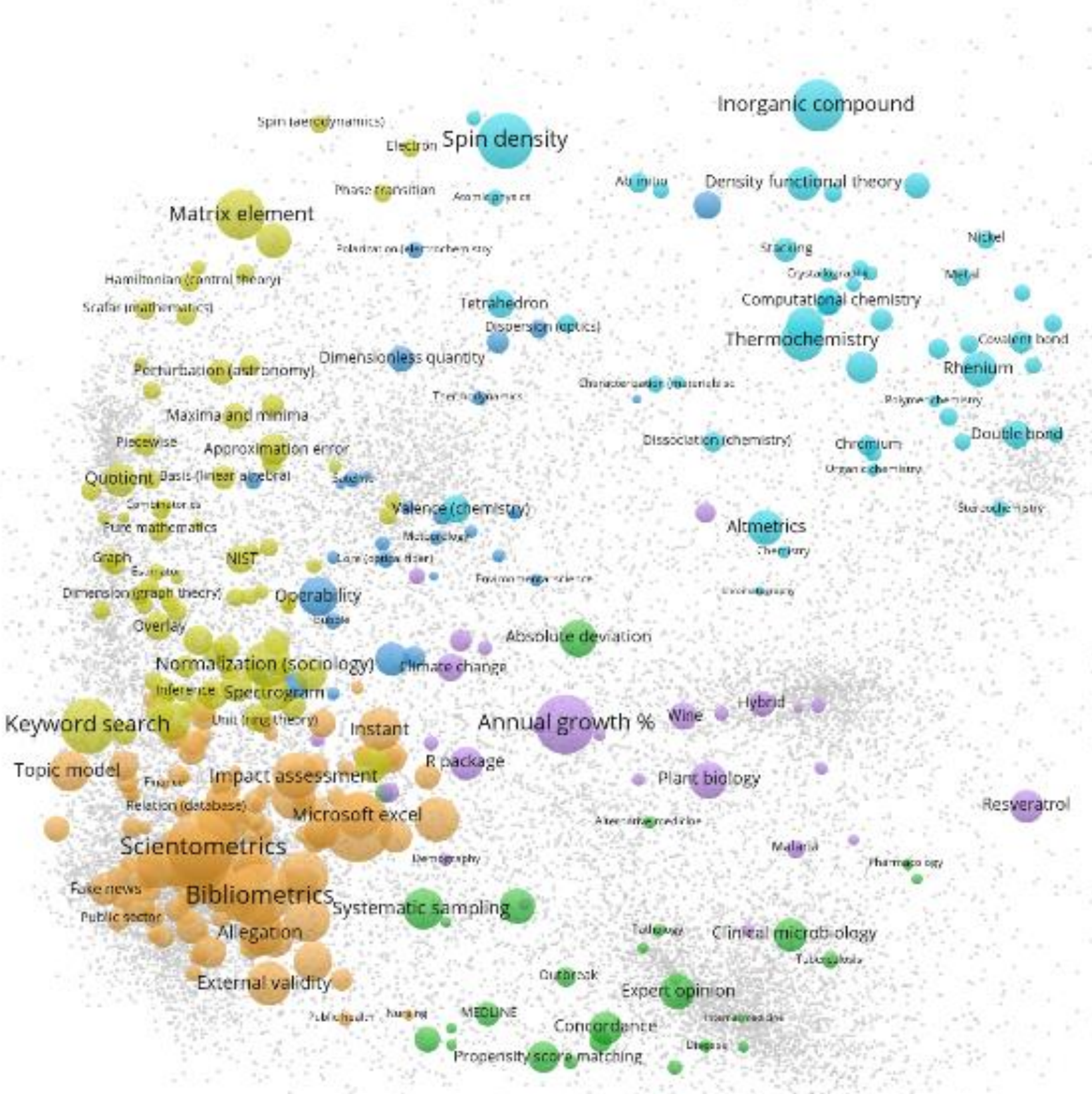

Figure 4. Normalized global overlay map for documents assigned to the OpenAlex author ID of the first author of this paper (RH)

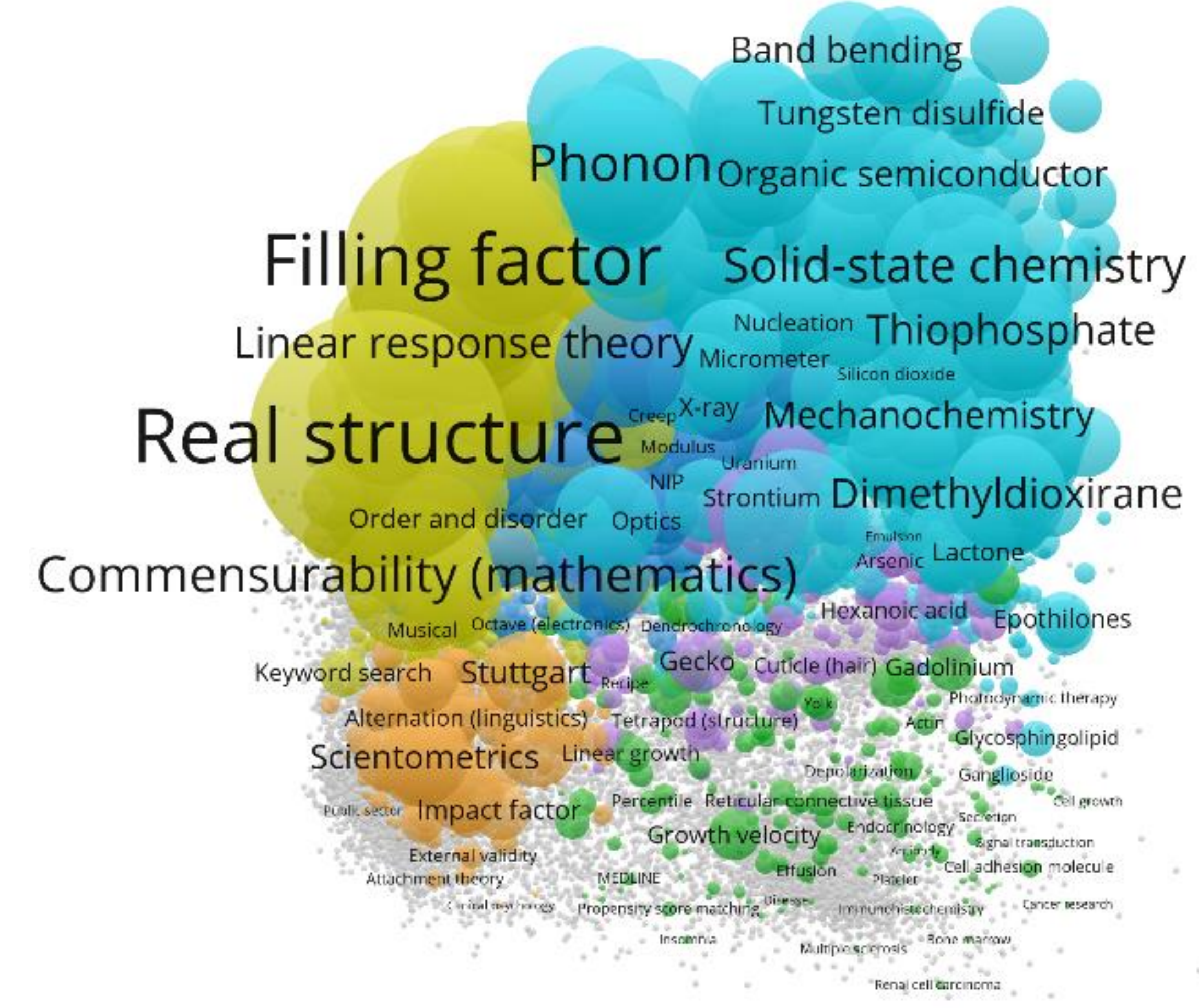

Figure 5. Normalized global overlay map for documents assigned to the OpenAlex institution ID of the Max Planck Institute for Solid State Research (MPI-FKF)

## 5. Discussion

The SSCT conceptualizes science as social systems consisting of linked communications. In this study, we have outlined a new approach of creating global overlay maps using OpenAlex with raw and normalized overlay data. These maps are empirical representations of social science systems: The research of the selected science unit is meaningfully visualized as linked (cited) formal communications (papers). In this study, we have provided six different base maps that readers can use to overlay their own data such as institutional or individual data. To exemplify the use of overlays with the base maps, we have provided two maps (data of one author and his research institute) using raw and normalized overlay data. The raw overlay maps presented here show that most concepts can be well associated with the unit's research areas and thus provide helpful guidance in interpretation of the units' research. The overlay maps can be used to explore the topics of both publication sets although one needs to be cautious when surprising concepts appear.

The overlay maps are suitable to explore the topics of different types of units, e.g., individuals, research groups, research institutes, or countries. Our general impression is that the raw overlay maps emphasize higher level (general) concepts more, and normalized overlay maps emphasize lower level (specialized) concepts more. This can be explained as follows: Since many more papers are assigned to higher level concepts, the higher level concepts are more visible on the raw overlay maps. Normalization of the overlay maps compares the focal unit's activity to the world's activity. Thereby, the higher level concepts loose prominence compared to lower level concepts. If one is interested in comparing different research units, normalized maps will be better suited than maps that are not normalized in most of the cases. Both maps (normalized and not normalized) provide different perspectives when gaining information about single research units: A more general perspective can be obtained using the not normalized maps while a more detailed perspective is provided by the normalized maps.
The main advantage of our approach based on OpenAlex data to produce overlay maps is that the underlying data can be used without any restrictions. Data for global base maps (covering the complete science system) and overlays can be downloaded without any costs. The data are freely available. What are the limitations of the approach? One of the main limitations concerns the concepts provided by OpenAlex. The visual inspection of many individual maps (the maps presented in this paper and several other maps) revealed that some concepts are erroneously assigned to papers. We also found out that some concepts are only partly correct; especially the additions in parentheses to labels were frequently erroneous or at least misleading. We hope that the reliability and validity of the process at OpenAlex to assign concepts to publications will be further improved in the near future. The announced new topics classification system[2] might provide improved alternative classifications of documents in OpenAlex.
Another limitation of this study concerns the used normalization method which may not be perfect. The method compares all concepts with each other irrespective of scientific field. Overlay maps using normalization with respect to scientific field of the concepts might produce different views. In the current normalization procedure, the number of papers in each concept is divided by the number of all papers with a concept on the same level of the unit using multiplicative counting. Alternatively, one could use the total number of papers of the unit that have a concept. Since the normalization method can present rarely occurring concepts very prominently because comparably few publications were assigned to this concept, users may introduce thresholds for the number of papers per concept that prevent concepts with too few papers to be displayed. A third limitation of our approach is related to the base maps that have been produced by us for recent years. Although these base maps are based on large publication sets, some concepts are located further away from most concepts. For generating some of the base maps' images, it was necessary to zoom into the maps and to cut the outliers. This limitation, however, concerns only the maps based on papers from comparably short time periods.

## 5. Conclusions

The visualization approach based on OpenAlex data that we introduced in this study can be used for any science unit. With the present study, we tried to develop a workflow by which the user is in a good position to produce meaningful overlay maps (without any costs or fees). We also assume that the approach introduced in this study could (should) be extended in future studies. We think, e.g., that it would be interesting to consider a broader set of overlay data. The current approach visualizes publication output (i.e., number of publications) on certain research topics. Node colors could easily be changed from field assignment to average publication years of the underlying documents. The number of publications published in reputable journals and the normalized citation impact of publications are other interesting options and could be well integrated in overlay maps, e.g., by node coloring. However, they are currently not available in OpenAlex.

[2]https://groups.google.com/g/openalex-users/c/2yE1jie_D3s/m/c3j9UYiLBgAJ?utm_medium=email&utm_source=footer

## Open science practices

The freely available open database OpenAlex was used so that readers can reproduce our results and use the base maps we provide. Our data can be obtained from https://openalex.org and our base maps can be downloaded from http://ivs.fkf.mpg.de/global_maps_OpenAlex (Haunschild & Bornmann, 2024). The procedures to create the base maps used free software tools (PostgreSQL, VOSviewer, and R). An example SQL for creating the base maps was included in the online resource connected to this paper (Haunschild & Bornmann, 2024). R scripts for post-processing of the SQL results and OpenAlex API requests are available upon request from the first author.

## Acknowledgments

The bibliometric data used in this paper are from a custom database of OpenAlex hosted at the Competence Network for Bibliometrics (see also https://bibliometrie.info/).

## Author contributions

Conceptualization: RH
Data curation: RH
Formal Analysis: RH, LB
Investigation: RH, LB
Methodology: RH, LB
Project administration: RH
Supervision: RH
Validation: RH, LB
Visualization: RH
Writing – original draft: RH, LB
Writing – review & editing: RH, LB

## Competing interests

The authors declare that they have no competing interests.